\documentclass{JAC2003}
\usepackage{graphicx}
\usepackage{booktabs}
\usepackage{array}
\usepackage{subfigure}
\usepackage{cite} 
\usepackage[usenames,dvipsnames]{xcolor}
\usepackage{hyperref}  
\hypersetup{bookmarks=true, pdfauthor={Pei Zhang},colorlinks=true,linkcolor=black,citecolor=blue,breaklinks=true}

\begin{document}
\title{THE TUNING SYSTEM FOR THE HIE-ISOLDE HIGH-BETA \\QUARTER WAVE RESONATOR\\[-.8\baselineskip]}

\author{P.~Zhang$^{1,}$\thanks{pei.zhang@cern.ch}, L.~Alberty$^{1}$, L.~Arnaudon$^{1}$, K.~Artoos$^{1}$, S.~Calatroni$^{1}$, O.~Capatina$^{1}$,\\ A.~D'Elia$^{1,2,3}$, Y.~Kadi$^{1}$, I.~Mondino$^{1}$, T.~Renaglia$^{1}$, D.~Valuch$^{1}$, W.~Venturini~Delsolaro$^{1}$\\
\mbox{$^1$CERN, Geneva, Switzerland}\\
\mbox{$^2$School of Physics and Astronomy, The University of Manchester, Manchester, U.K.}\\
\mbox{$^3$The Cockcroft Institute of Accelerator Science and Technology, Daresbury, U.K.}}

\maketitle


\begin{abstract}
A new linac using superconducting quarter-wave resonators (QWR) is under construction at CERN in the framework of the HIE-ISOLDE project. The QWRs are made of niobium sputtered on a bulk copper substrate. The working frequency at 4.5~K is 101.28~MHz and they will provide 6~MV/m accelerating gradient on the beam axis with a total maximum power dissipation of 10~W on cavity walls. A tuning system is required in order to both minimize the forward power variation in beam operation and to compensate the unavoidable uncertainties in the frequency shift during the cool-down process. The tuning system has to fulf{}ill a complex combination of RF, structural and thermal requirements. The paper presents the functional specif{}ications and details the tuning system RF and mechanical design and simulations. The results of the tests performed on a prototype system are discussed and the industrialization strategy is presented in view of f{}inal production.
\end{abstract}

\section{Introduction}
The HIE-ISOLDE \cite{hie-1} project is a major upgrade of the existing ISOLDE radioactive beam facility at CERN. The main focus is to boost the beam energy from 3~MeV/u to 10~MeV/u by replacing the current normal conducting linac with superconducting quarter wave resonators (QWRs). They will make use of niobium sputtered on copper substrate technology \cite{srf13-1}. The QWR will have a working frequency of 101.28~MHz at 4.5~K providing an accelerating gradient of 6~MV/m on beam axis with a maximum of 10~W power dissipation. Two types of QWRs, low-$\beta$ and high-$\beta$, will be installed to cover the entire energy range. Since the linac upgrade will start from the high energy section, the R\&D ef{}fort has been focused on the high-$\beta$ QWRs \cite{srf13-2,srf13-3,srf13-5}. 

The resonant frequency of each QWR varies inevitably due to mechanical tolerances and uncertainties during the cool-down process. We decoupled these two ef{}fects \cite{hie-2,hie-note-3,hie-note-4}, thus the tuning system only needs to compensate the variability of the frequency shift during the cool-down process. Due to insuf{}f{}icient statistics on the frequency shift at the beginning, a rather large coarse range for the tuning system was chosen. Measurements from the last two years have pinned down the uncertainty of the frequency shift. Therefore it has been possible to go for a simplif{}ied tuning system which will also lower the production cost. In addition, the system must also provide a f{}ine frequency tuning in order to keep the cavity on resonance hence minimizing the forward power variation during beam operations. 

This paper f{}irst shows the study on the frequency shift during the cool-down process based on previous measurements. Then it reviews the current tuning plate with two dif{}ferent assemblies. Aiming at a lower sensitivity, the study of a simplif{}ied tuning plate is described as well as its impact on the main cavity RF parameters. Finally the mechanical implementation and measurement results of two dif{}ferent simplif{}ied plates are presented.

\section{Frequency Shift during the Cool-down Process}
\begin{figure}[h]\centering
\includegraphics[width=0.45\textwidth]{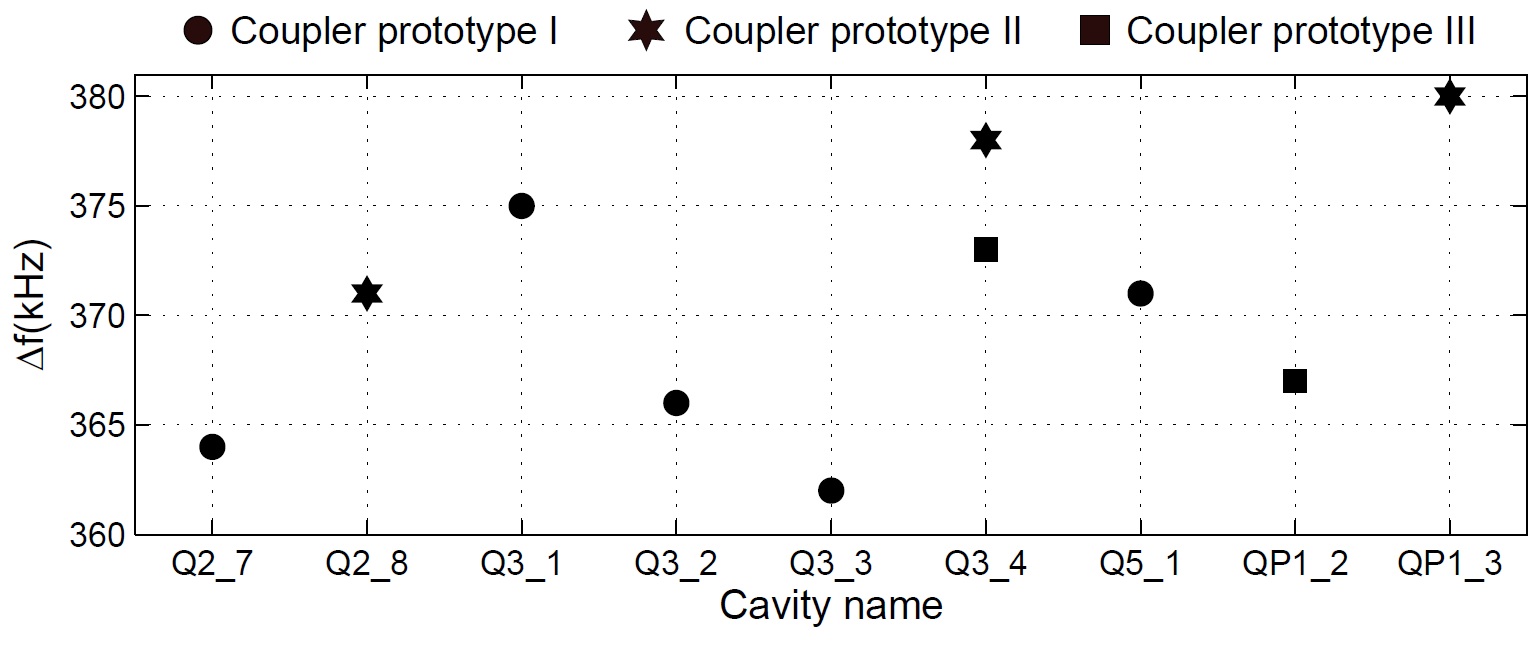}
\caption{Frequency shift during the cool-down process.}
\label{freq-scale}
\end{figure}

The resonant frequency of the cavity shifts during cool-down process \cite{hie-note-1}, and this shift varies amongst cavities and coatings. Fig.~\ref{freq-scale} shows the results from our previous measurements on various QWRs using three dif{}ferent couplers. It suggests a peak-to-peak frequency shift uncertainty of 18~kHz. This has to be covered by the coarse range of the tuning system. To stay in a safer limit, we doubled this value to 36~kHz for the coarse range, which is comparable to TRIUMF (33~kHz) \cite{triumf-1} and Legnaro (25~kHz) \cite{legnaro-1}.

\section{The original Tuning Plate}
\begin{figure}[h]\centering
\includegraphics[width=0.47\textwidth]{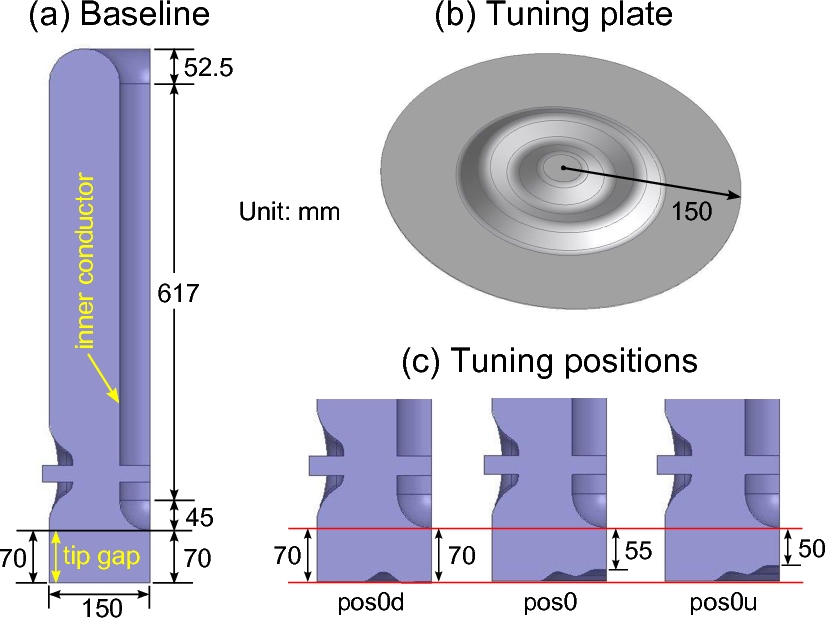}
\caption{Baseline QWR with tuning plate assembled upwards.}
\label{qwr-baseline}
\end{figure}

The original tuning system is conceptually similar to TRIUMF's and has been described in \cite{hie-2}. An oilcan-shaped diaphragm of copper-beryllium (CuBe) was hydroformed and then coated with niobium. Fig.~\ref{qwr-baseline}(b) shows schematically the tuning plate in mid-range position ``pos0''. This plate was designed to be assembled upwards on the bottom of the baseline QWR (Fig.~\ref{qwr-baseline}(a)) as shown in Fig.~\ref{qwr-baseline}(c). The tuning range is 20~mm consisting 5~mm upward and 15~mm downward movement from ``pos0''. The coarse range is calculated to be 220~kHz giving an average tuning sensitivity of 11~kHz/mm. During beam operations, microphonics, helium pressure variation and Lorentz force will constantly change the cavity resonant frequency which need to be tuned in a ``f{}ine'' manner. The LLRF requires 0.5~Hz per tuning step \cite{hie-note-2}, which can be translated into a mechanical step of 45~nm. This makes the mechanical control very challenging. 

\begin{figure}[h]\centering
\includegraphics[width=0.45\textwidth]{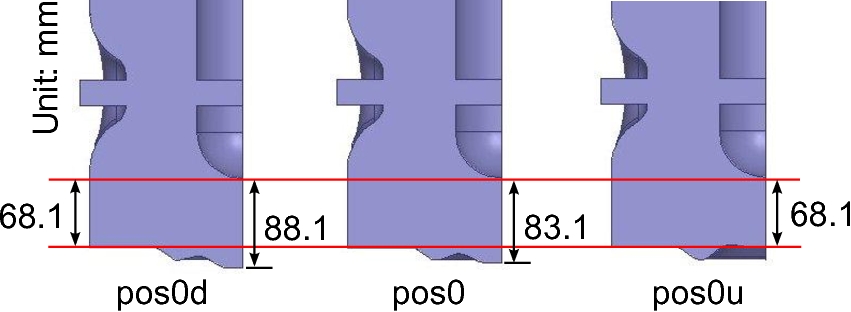}
\caption{Baseline QWR with tuning plate assembled downwards. The inner conductor has been elongated by 1.9~mm to tune back the nominal frequency.}
\label{qwr-base-down}
\end{figure}

A possible solution is to f{}lip the existing tuning plate to assemble it downwards as shown in Fig.~\ref{qwr-base-down}. By doing that, the ef{}fective tip gap is enlarged from 55~mm to 83.1~mm for ``pos0''. The inner conductor of the cavity has been elongated by 1.9~mm from the baseline (Fig.~\ref{qwr-baseline}(a)) in order to tune back the nominal frequency. The simulation results for three tuning positions are listed in Table~\ref{table-baseline-fancy}. Simulations suggest a reduced sensitivity of 6.85~kHz/mm and a smaller coarse range of 137~kHz accordingly. In this case, a mechanical step of 73~nm would be suf{}f{}icient to fulf{}il the 0.5~Hz/step LLRF requirement, making the mechanical control less challenging. However this tuning plate is highly costly and the 137~kHz coarse range is not really necessary. A simplif{}ied, low-cost tuning plate is therefore preferable.

\begin{table}[h]\center
\caption{Tuning Sensitivity and Coarse Range of the Original Plate Assembled on the Baseline QWR both Upwards and Downwards. The results are from simulations at room temperature in air.}
\label{table-baseline-fancy}
\begin{tabular}{c|c|c}
\hline
& freq. (upwards) & freq. (downwards) \\
\hline
pos0d & 100.933~MHz & 100.937~MHz\\
\hline
pos0 & 100.804~MHz & 100.911~MHz\\
\hline
pos0u & 100.713~MHz & 100.800~MHz\\
\hline
\textbf{Coarse range} & \textbf{220~kHz} & \textbf{137~kHz}\\
\hline
\textbf{Sensitivity} & \textbf{11~kHz/mm} & \textbf{6.85~kHz/mm}\\
\hline
\end{tabular}
\end{table}

\section{The New Simplif{}ied Tuning Plate}

\begin{figure}[h]\centering
\includegraphics[width=0.47\textwidth]{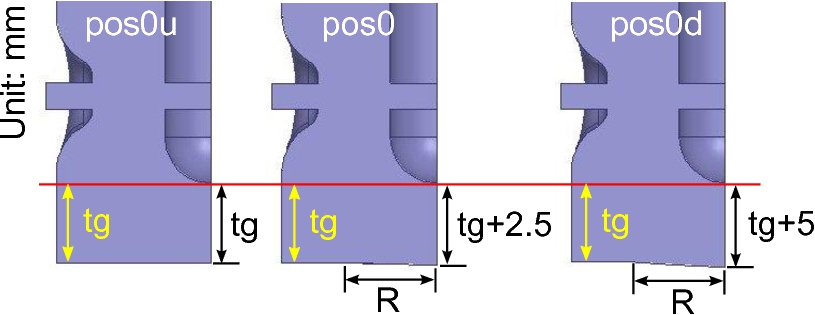}
\caption{Simplif{}ied tuning plate; $tg$ stands for tip gap and $R$ is the radius of the deformation region.}
\label{qwr-simple}
\end{figure}

\begin{figure}[h]\centering
\includegraphics[width=0.47\textwidth]{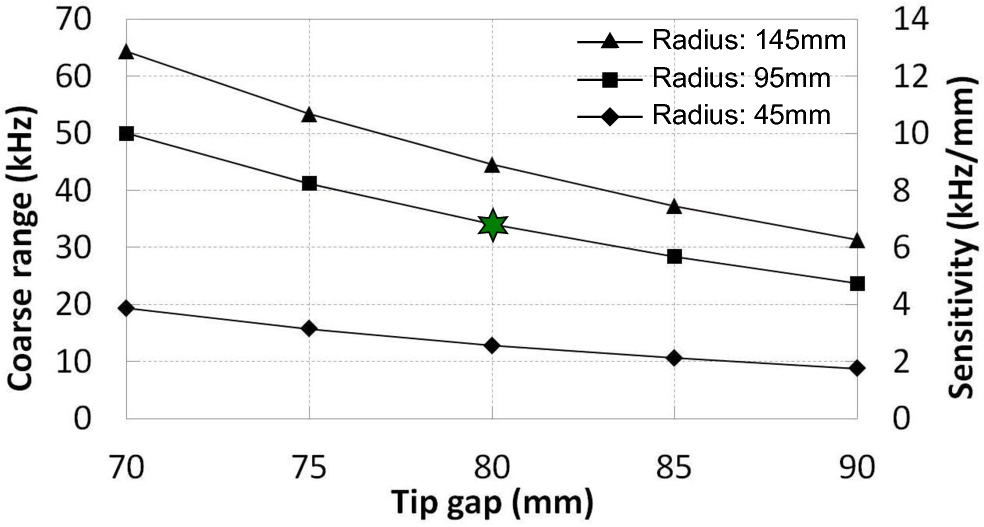}
\caption{Coarse range and tuning sensitivity of the simplif{}ied plates with dif{}ferent deformation radius.}
\label{qwr-simple-range}
\end{figure}

Starting from a simple f{}lat plate, the tuning is made by deforming the plate from the center. Fig.~\ref{qwr-simple} shows the simplif{}ied plate along with the cavity at dif{}ferent tuning positions. The tuning is realized by pushing/pulling the plate by a maximum of 2.5~mm. Fig.~\ref{qwr-simple-range} shows the frequency tuning varies with the radius of the deformable area. It can be clearly seen that $R$=95~mm at $tg$=80~mm (the green hexagram in Fig.~\ref{qwr-simple-range}) would give a reasonable choice with a coarse range of 34~kHz and a sensitivity of 6.8~kHz/mm. 

\begin{figure}[h]\centering
\includegraphics[width=0.45\textwidth]{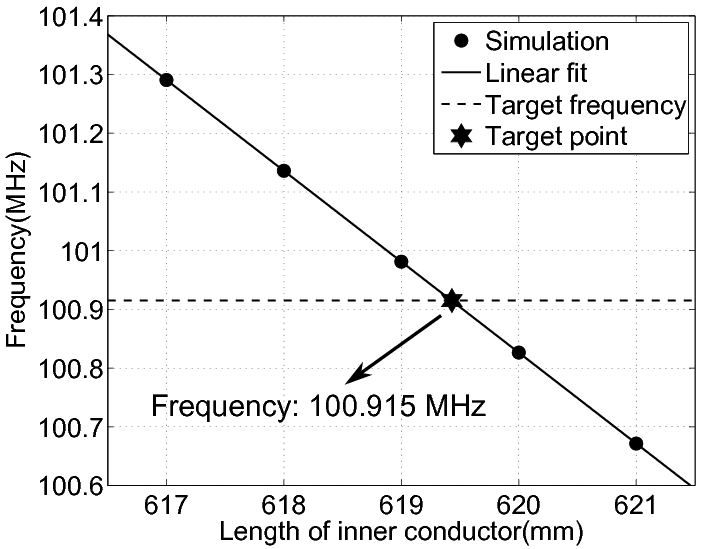}
\caption{The tuning of the cavity frequency by changing the inner conductor length.}
\label{qwr-sweepAntL}
\end{figure}

Compared to the baseline design in Fig.~\ref{qwr-baseline}, the tip gap has been increased by 10~mm. This will inevitably increase the cavity frequency by several hundred kHz, thus the cavity needs to be retuned. We assume \cite{hie-note-1} a frequency shift of +365~kHz to characterize the changes from room temperature in air to 4.5~K in vacuum. Thus 
\begin{equation}
f^{eigenmode}_{\mathrm{simulation}} = 101.28 - 0.365 = 100.915~\mathrm{MHz}.
\label{eq:freq}
\end{equation}
As the cavity frequency is very sensitive to the change of the inner conductor length, we decided to modify this parameter to tune back the frequency in order to minimize the changes on the overall cavity design. Fig.~\ref{qwr-sweepAntL} shows how the frequency varies with the inner conductor length. After a linear f{}it of the simulation results, the f{}itted line intersects with the target frequency line giving the required inner conductor length. It suggests an elongation of the inner conductor by 2.5~mm.

\begin{figure}[h]\centering
\includegraphics[width=0.45\textwidth]{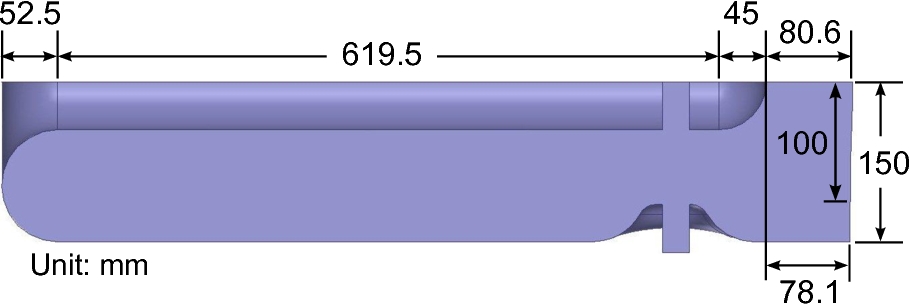}
\caption{Current cavity with the simplif{}ied tuning plate at mid-range position.}
\label{qwr-new}
\end{figure}

The simplif{}ied tuning plate will have a deformable area with a radius of 100~mm, a mid-range position at 2.5~mm deformed downwards with a total movable range of 5~mm ($\pm$2.5~mm around the mid-range position). Finally a f{}ine cavity tuning gave a tip gap of 78.1~mm. Fig.~\ref{qwr-new} shows the f{}inal cavity geometry.

Since our goal is to minimize the changes to the overall cavity design, we examined the main RF parameters of each design. These results are listed in Table~\ref{table-rf-par}. Due to an increased tip gap in the new design, the E$_p$/E$_{acc}$ decreases by 10\% from the baseline, which may reduce the occurrence of f{}ield emission. The magnetic f{}ield on the plate is also lowered, which cuts down the power dissipation on the plate by 49\% from the baseline. In particular, the magnetic f{}ield at the contact between the plate and the cavity outer wall is lowered by 10\%. This would reduce the loss due to contact resistance. The changes to other RF parameters are negligible.
\begin{table}[ht]\center
\caption{The Main Cavity RF Parameters with Dif{}ferent Tuning Plate from Simulations at Room Temperature in Air.}
\label{table-rf-par}
\begin{tabular}{l|c|c|c}
\hline
& \textbf{Up.} & \textbf{Down.} & \textbf{New} \\
\hline
$f_{simulation}^{eigenmode}$ [MHz] & 100.804 & \textcolor{Black}{100.911} & \textcolor{Black}{100.914} \\
\hline
$\beta_{optimum}$ [\%] & 10.86 & \textcolor{Black}{10.87} & \textcolor{Black}{10.88}\\
\hline
TTF at $\beta_{optimum}$ & 0.90 & \textcolor{Black}{0.90} & \textcolor{Black}{0.90}\\
\hline
R/Q [$\Omega$] (incl. TTF) & 554 & \textcolor{Black}{556} & \textcolor{Black}{556}\\
\hline
E$_{p}$/E$_{acc}$ & \textbf{5.5} & \textbf{5.0} & \textbf{5.0} \\
\hline
H$_{p}$/E$_{acc}$ [G/(MV/m)] & 95.4 & \textcolor{Black}{95.3} & \textcolor{Black}{95.3}\\
\hline
U/E$_{acc}^2$ [mJ/(MV/m)$^2$] & 208 & \textcolor{Black}{207} & \textcolor{Black}{207} \\
\hline
G=R$_s$Q [$\Omega$] & 30.7 & \textcolor{Black}{30.8} & \textcolor{Black}{30.8}\\
\hline
P$_{diss}$ [W] in the cavity & 7.7 & \textcolor{Black}{7.7} & \textcolor{Black}{7.7}\\
\hline
P$_{diss}$ [W] on the plate & \textbf{0.0035} & \textbf{0.0021} & \textbf{0.0018} \\
\hline\hline
Conductor $L$ [mm] & 617 & \textcolor{Black}{618.9} & \textcolor{Black}{619.5} \\
\hline
Tip gap [mm] & 70 & \textcolor{Black}{68.1} & \textcolor{Black}{78.1}\\
\hline\hline
Coarse range [kHz] & 220 & 137 & 37 \\
\hline
Sensitivity [kHz/mm] & 11.0 & 6.9 & 7.4 \\
\hline
\end{tabular}
\end{table} 

Allowing for mechanical tolerances of 0.1~mm, a detuning of 28~kHz from the nominal cavity frequency can be expected in the worst case. For this reason, the tip gap value has been chosen to be a free parameter in order to compensate the mechanical-tolerance induced frequency detuning \cite{hie-note-3}. Simulations suggest that a change of approximately $\pm$1.5~mm to the tip gap is necessary to recover the $\mp$28~kHz frequency detuning. The impact of this possible tip gap variations on the coarse range and tuning sensitivity has been studied by simulations and listed in Table~\ref{table-tolerance}. The 1.5~mm variation of the tip gap will shift the coarse range by 2~kHz and the sensitivity by 0.2~kHz/mm. These are small and tolerable.
\begin{table}[h]\center
\caption{The Coarse Range and Tuning Sensitivity at Dif{}ferent Tip Gap with $R$=100~mm Simplif{}ied Plate.}
\label{table-tolerance}
\begin{tabular}{l|c|c|c}
\hline
& \multicolumn{3}{c}{Tip gap (mm)} \\
\hline
& 78.1-1.5 & 78.1 & 78.1+1.5 \\
\hline
Coarse range (kHz) &  38 & 37 & 35 \\
\hline
Sensitivity (kHz/mm) & 7.6 & 7.4 & 7.0 \\
\hline
\end{tabular}
\end{table}

\section{The Mechanical Design of the Simplif{}ied Tuning System}
The mechanical deformation of the tuning plate is created by a stepper motor in microstep mode. It has 8000 steps per turn which is transformed to a linear motion by a screw at room temperature with 1~mm advance per turn. The theoretical smallest motor step of 0.125~$\mu$m was conf{}irmed by measurements. The precision is however reduced to the micron range by friction, depending on the motion range and load. The hysteresis is 1--2~$\mu$m when changing moving directions. 

The linear motion is transmitted by a pulling rod with two hinges to a mechanical lever changing the motion direction at the bottom of the cavity. The lever further divides the mechanical motion by a factor of 3 in order to reach a theoretical resolution of 40~nm.
  
As mentioned above, the total deformation range of the tuning plate is 5~mm, pulling down from the f{}lat position. Given the space available for the mechanical lever, this range results in a lever angle up to 10$^{\circ}$. This linear and angular range along with the force required to deform the tuning plate complicates the use of a precise f{}lexural and hence frictionless guidance. For the f{}irst test an adjusted frictional guidance was used with material combination of copper on stainless steel in order to reduce the risk of stick slip and cold welds at 4.5~K in vacuum. A knife edge pivot can also be considered to further reduce the friction.

The change from an oilcan-shaped diaphragm to a f{}lat tuning plate introduced difficulties for mechanical and thermal design. A f{}lat plate design implies a non-linear stif{}f behavior. In addition, stresses and the force required for a 5~mm moving range increase signif{}icantly. The f{}irst prototype was produced in 0.3~mm thick CuBe C17140, which has high elastic limit and acceptable thermal conductivity at 4.5~K. This thin plate was clamped against the cavity between 300~mm and 320~mm in diameter. The radius of the deformable area was limited to 100~mm by a limiting Cu~OFE ring under the plate. This ring might also help to conduct the dissipated RF power on the plate. The second prototype was produced in Cu~OFE. Starting from a 5~mm thick Cu~OFE disk, the deformable area with a radius of 100~mm was machined to 0.3~mm. This plate will be plastif{}ied locally before reaching the maximum 5~mm deformation. This requires higher forces but have a significantly better thermal behavior compared to the CuBe plate. It therefore has suf{}f{}icient thermal margin to keep the coated niobium layer to remain superconducting during beam operations.

Both plates have been coated with niobium in order to minimize losses. Surface treatment prior to coating was performed in the same way as for the cavities \cite{srf13-2}, which removed approximately 20~$\mu$m. However the assessment of our surface treatment as optimal method for CuBe surface preparation has still to be performed. The f{}inal preparation step is a low pressure, dust-free, demineralized water rinsing. The plates are then packed in dust-free polymer bags for transfer to the coating laboratory. Niobium coating was performed by sputtering in a dual planar magnetron system, dif{}ferent from the one used for cavities \cite{srf13-2,srf13-3}. Niobium layers of approximately 1.5~$\mu$m thick were coated on both plates with the RRR exceeding 15. Final production step was again the dust-free water rinsing prior to assembling on the cavity. The quality factors $Q_0$ in excess of 2$\times$10$^9$ ($R_s$$<$15~$n\Omega$) at low f{}ield have been measured for the same reference baseline cavity using both tuning plates, indicating that the coating of the plate is not an intrinsic limiting factor of the performance. Adhesion of the coating is generally satisfactory, except sometimes peel-of{}f at the edges of the plate was observed. This is irrelevant from the RF point of view, and could be traced to contaminations by the polymer bags which must be used for packaging to prevent surface contamination from dust.

\section{Measurement results of the Simplif{}ied Tuning System}
\begin{figure}[h]\centering
\includegraphics[width=0.45\textwidth]{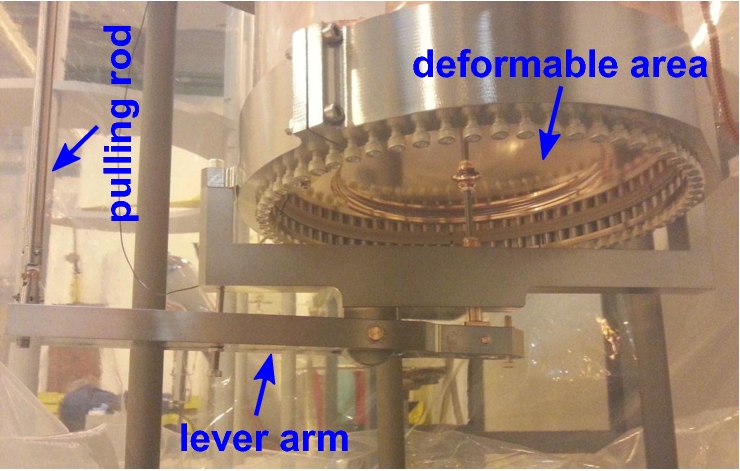}
\caption{Measurement setup for the tuning system test (photo taken outside the cryostat).}
\label{CuOFE-setup}
\end{figure}

The tuning plate was installed on a coated baseline cavity with 80~mm tip gap and mechanically controlled by the above stated stepper motor. Fig.~\ref{CuOFE-setup} shows the measurement setup. Measurements of coarse range and sensitivity were subsequently conducted at 4.5~K for both CuBe and Cu~OFE plate.

\begin{figure}[h]
\subfigure[Frequency tuning (CuBe)]{
\includegraphics[width=0.44\textwidth]{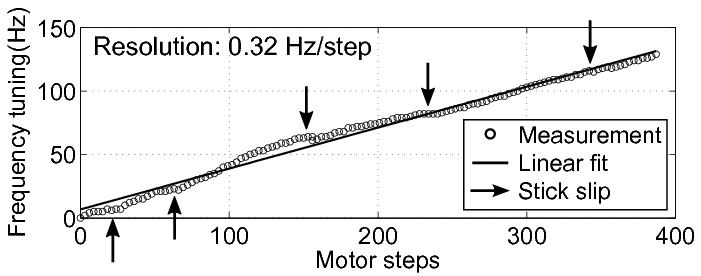}
\label{CuBe-tuning}
}
\subfigure[Frequency error (CuBe)]{
\includegraphics[width=0.48\textwidth]{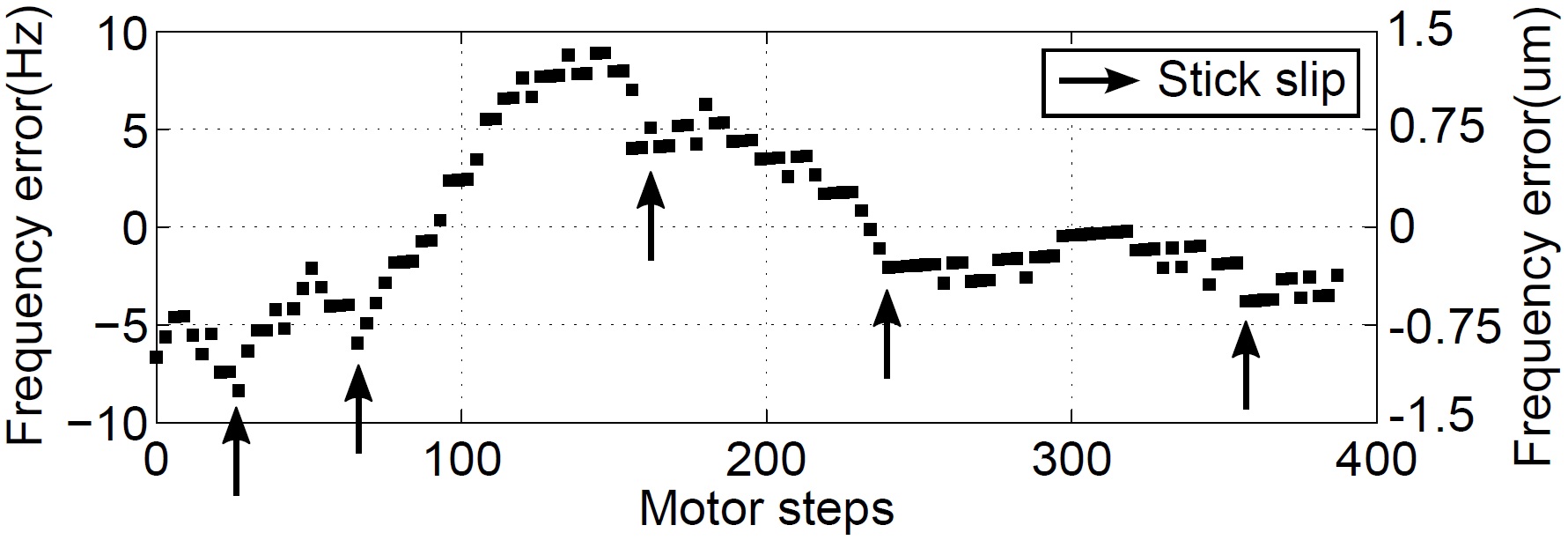}
\label{CuBe-error}
}
\caption{Measurement results of the CuBe plate.}
\label{CuBe}
\end{figure}

Fig.~\ref{CuBe-tuning} shows an example of the measurement results for the CuBe plate. A total frequency tuning of 130~Hz was tested in 390 motor steps with 3 steps per movement. The frequency tuning followed the motor step changes linearly with a resolution of 0.32~Hz/step. Stick slip were observed several times during the tuning. This results in deviations in the order of 5~Hz between the measured frequency and the linear f{}it as shown in Fig.~\ref{CuBe-error}. Such frequency errors correspond mechanically to the micron level as can be expected in the best case for a frictional system. The coarse range for the CuBe plate was measured to be 27~kHz at 4.5~K.

\begin{figure}[h]
\subfigure[Frequency tuning (Cu~OFE)]{
\includegraphics[width=0.45\textwidth]{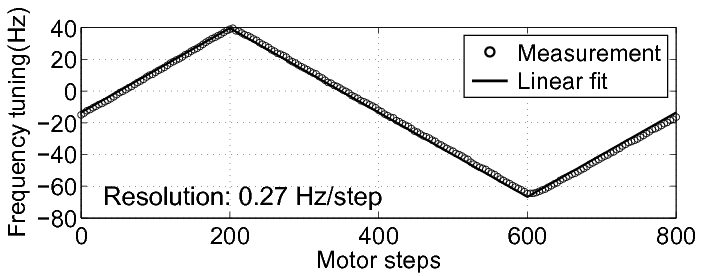}
\label{CuOFE-tuning}
}
\subfigure[Frequency error (Cu~OFE)]{
\includegraphics[width=0.48\textwidth]{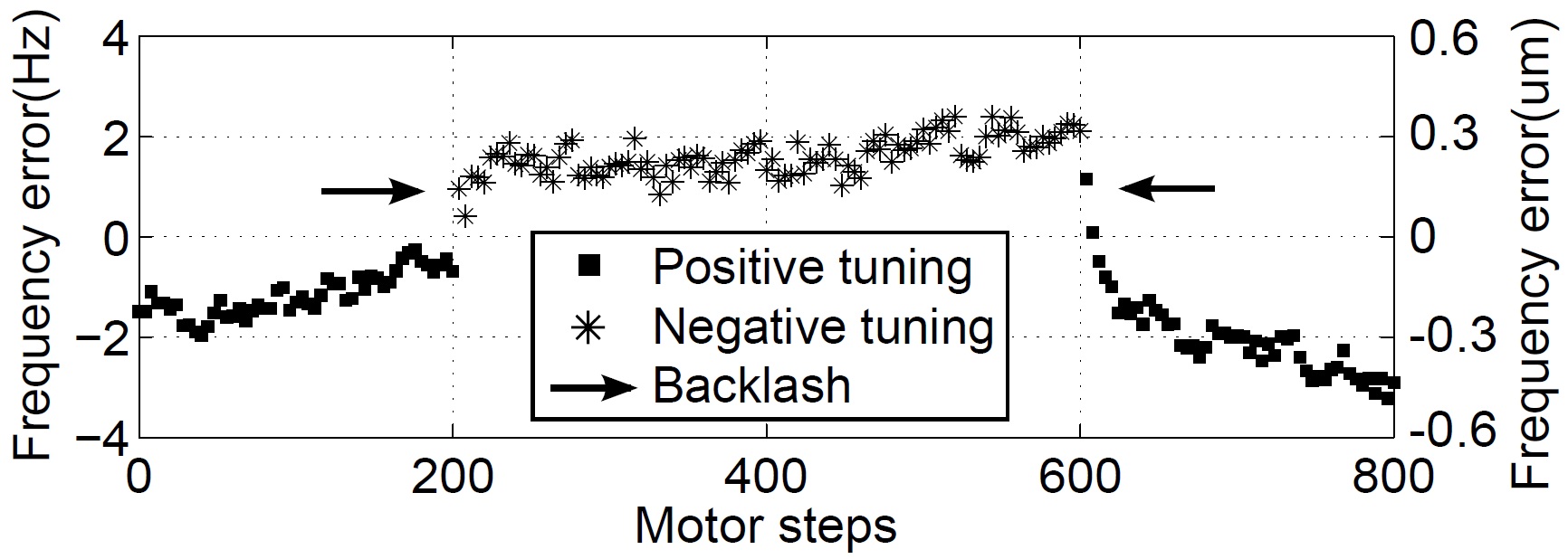}
\label{CuOFE-error}
}
\caption{Measurement results of the Cu~OFE plate.}
\label{CuOFE}
\end{figure}

One example of the measurement results for the Cu~OFE plate are shown in Fig.~\ref{CuOFE-tuning}. The motor was moved in both directions in 40~steps each time to test both positive and negative tuning. A similar stick slip behavior was observed and a resolution of
0.27~Hz/step was measured. As shown in Fig.~\ref{CuOFE-error}, frequency errors are in the same order with the CuBe plate. The coarse range of the Cu~OFE plate was measured to be 24~kHz, smaller than that of the CuBe plate. For both plates, backlash was observed when changing moving directions as shown in Fig. 5(b) for the Cu~OFE plate. 

Coarse range was measured to be smaller than that of the design value for both prototypes. This was because we set some restrictions on the lever arm to limit the deformation. Later measurements at warm have conf{}irmed the desired coarse range of 37~kHz for the Cu~OFE plate. This is perfect consistency between simulations and measurements. Compared to CuBe plate, the Cu~OFE plate exhibited a more stable behavior during the test. 

\section{Final Remarks}
The frequency shift during the cool-down process of the HIE-ISOLDE high-$\beta$ quarter-wave resonators has been characterized based-on measurements from the last two years. A much smaller coarse range of 36~kHz along with less sensitivity was then proposed. Therefore a simplif{}ication of the original tuning system was favored in order to reduce the production cost. Two prototypes of the simplif{}ied tuning plates have been designed, manufactured and coated with niobium at CERN. They were subsequently installed on the cavity and tested at 4.5~K. Both plates were able to tune the cavity frequency linearly with the stepper motor control. The f{}ield and power requirements for HIE-ISOLDE were reached for both plates. The resolutions of both plates were measured to be much better than 0.5~Hz/step required by LLRF. However the measured coarse ranges at cold were lower than the design value for both plates which was due to restrictions set on the lever arm. The measurement results f{}it simulations perfectly for both coarse range and sensitivity. Given the large margin left for the resolution, we plan to enlarge the deformable area of the plate to increase the coarse range and lower the forces applied on the plate.    

\section{Acknowledgements}
We thank M.~Therasse for his help in preparing the measurements. This work has been supported partly by a Marie Curie Early Initial Training Network Fellowship of the European Community's 7th Programme under contract number PITN-GA-2010-264330-CATHI.


\end{document}